\documentclass[12pt]{elsarticle}

\usepackage{amsmath,amsgen,amstext,amssymb}
\usepackage{amsbsy,amsopn,amsfonts}

\begin{document}

\sloppy

\begin{frontmatter}

\title{Block Advertisement Protocol}

\author{Danil Nemirovsky}

\address{danil.nemirovsky@gmail.com}

\begin{abstract}
Bitcoin, a decentralized cryptocurrency, has attracted a lot of attention from academia, financial service industry and enthusiasts. The trade-off between transaction confirmation throughput and centralization of hash power do not allow Bitcoin to perform at the same level as modern payment systems. Block Advertisement Protocol is proposed as a step to resolve this issue. The protocol allows block mining and block relaying to happen in parallel. The protocol dictates a miner to advertise the block it is going to mine allowing other miners to collect all the transactions in advance. When a block in mined, only header is relayed since most of data about the block is already known to all miners.
\end{abstract}

\end{frontmatter}

\section{Introduction}

When Satoshi Nakamoto has introduced Bitcoin in his classic article~\cite{nakamoto2009bitcoin}, Nakamoto has not anticipated the problem which the bound on block size can cause. Block size, currently upperbounded by 1 megabyte, has become a trade-off between two essential properties of Bitcoin network. On one hand, keeping block size as it is now does not allow Bitcoin to be any close to transaction throughput required for retail usage and demonstrated by VISA and Mastercard. On the other hand, big block size leads to concentration of hash power and may compromise the whole idea of Bitcoin as a decentralized cryptocurrency. 

Another important argument against block size increase is unfair advantage gained by a miner lucky enough to mine the next valid block that will be accepted by the network with high probability. The miner is able to start next block mining immediately while other miners firstly have to download the block. If the size of a block is large, then to propagate it through the network may take significant time. Other miners waste their computation power since while the some miners may have accepted the block, they are mining a block that will not be accepted by the network. Ethereum partially resolves the problem including such blocks, called uncles, into blockchain on a later stage allowing unlucky nodes to recover a significant fraction of the value lost otherwise.

To make such an unfair advantage less prominent, we may allow block propagation to be done in parallel with mining of the block. To achive it we propose Block Advertisement Protocol.

The rest of the paper is organized as follows. Section~\ref{current_protocol} gives the description of current Bitcoin protocal. We introduce Block Advertisement Protocol in section~\ref{advertisement_protocol}. And the article is concluded in section~\ref{conclusion}.

\section{Current protocol}
\label{current_protocol}

When a miner is started, it downloads the whole blockchain from peers. When the blockchain has been downloaded, the miner is able to start transaction confirmation process, namely it can mine new blocks.

Each miner relays transactions through the network which it receives either directly from a bitcoin spender or from other miners.

Miner includes coinbase transaction into the block effectively fixing miner coinbase address. Coinbase transaction and other transactions are organized into a Merkle tree.

Miner prepares block header including version, previous block hash, Merkle tree root, block creation timestamp, difficulty target and nonce, and calculates hash. Miner iterates through nonce space checking if a calculated hash meets the difficulty target. If the nonce space is exhausted, the miner can change $extraNonce$ in the coinbase transaction, include more transations into Merkle tree and/or update block creation timestamp.

When the required nonce is found and the block has hash, meeting the difficulty target, the block is sent to peers.

Since typical transaction size is $0.5$kB and block size is bounded by $1$ megabytes, we may have about $2000$ transactions in a block.

Current Bitcoin protocol does not incentivize miners to relay a transaction, in particular a transaction having high fee, since miner prefers to include the transactions into the block it mines instead of giving a chance to other miners. As a result of such miner behaviour, it can take longer than necessary to confirm some transactions.

\section{Block Advertisement Protocol}
\label{advertisement_protocol}

We propose to adopt Block Advertisement Protocol which allows to reduce latency of mined block propagation. The protocol assumes that transaction malleabality issue is solved. The protocol works as follows.

Miner downloads the whole blockchain to be ready to mine a block on the most fresh transaction history shared by the network.

When a miner $M$ has collected a number of transactions and is ready to start mining a block $B$, it advertises the content of block $B$ to other miners. Namely, miner $M$ shares an advert $A_c$ containing miner coinbase address $c$, ordered list of transaction hashes $L$ to be included into block $B$ and previous block hash $h$. Advert $A_c$ is relayed over the miner network to every miner and each miner keeps a mapping from miner coinbase address $c$ to miner advert $A_c$.

After that miner $M$ starts mining block $B$ keeping the list of transactions unchanged. The only fields it is allowed to change are $extraNonce$ in coinbase transaction, block creation timestamp and nonce.

While miner $M$ mines block $B$, other miners have a chance to download the transactions from advertised list $L$. They even can request a specific transaction since they have transaction hashes. Miner $M$ has to share all the transactions. It eliminates the possibility to keep a transaction in secret. By the time the block is mined, other miners may have all the transactions downloaded and validated againts previous block.

When block $B$ is mined, instead of relaying a whole block, miner $M$ shares block seed $S_B$ which contains coinbase address, coinbase transaction and block header $H$. Using coinbase transaction and transaction list $L$ every miner is able to reconstruct block $B$. If miner $M$ has indeed mined the block it had advertised, the Merkle root of the reconstructed block will match the Merkle root in header $H$. 

Other miners will accept the block from miner $M$, in other words will consider the block valid, if its coinbase address matches the one stored in mapping, namely $c$, it contains previous block hash, its hash meets difficulty target requirement, it includes transactions from the transaction list $L$ advertised in $A_c$.

Miners will accept only one advert from $M$ per previous block hash $h$ to prevent advert spamming from one miner.

Since miner $M$ has advertised block $B$ to be mined, it cannot add arriving transaction into the block. Hence, miner $M$ will validate new transactions against block $B$ preparing the foundation for the next block. If block $B$ is accepted by the network, miner $M$ may already have the list of transaction for the next block to advertise. In this case miner $M$ can issue next advert right after block $B$ is mined. If miner $M$ is not lucky and it received a block $B'$ mined by another miner, miner $M$ has to validate transactions against block $B'$ and issue another advert containing hash of block $B'$.

A miner may adopt the tactic to share an advert just after the corresponding block is mined. Although in this case the miner can include more transaction into the block when they come to the miner, the probability that the block is accepted by the network decreaces since other miners cannot actively request transactions that could be advertised in advance but were not.

If we assume that block contains only list of transaction without other auxiliary information like Merkle tree and header and typical transaction size is $0.5$kB, advert will be about $15$ times smaller since it contains only transaction hashes.

\section{Conclusion}
\label{conclusion}

The proposed Block Advertisement Protocol allows a miner to mine and relay a block in parallel. The miner advertises the block it is going to mine to give the chance other miners to collect all the transactions in advance. When a block is mined, only header is relayed since most of data about the block is already known to all miners. Hence, the information about mined block can be  spread in the network quicker reducing the time when other miners can start their mining process based on the new valid block and, therefore, the unfair advantage gained by the lucky miner. The protocol is a step to make Bitcoin and other blockchain as performant as modern payment systems.

\end{document}